\documentstyle[floats,aps]{revtex}
\begin{document}
\input epsf

\font\twelvemib=cmmib10 scaled 1200
\font\elevenmib=cmmib10 scaled 1095
\font\tenmib=cmmib10
\font\eightmib=cmmib10 scaled 800
\font\sixmib=cmmib10 scaled 667
\skewchar\elevenmib='177
\newfam\mibfam
\def\mib{\fam\mibfam\tenmib}
\textfont\mibfam=\tenmib
\scriptfont\mibfam=\eightmib
\scriptscriptfont\mibfam=\sixmib
\mathchardef\alpha="710B
\mathchardef\beta="710C
\mathchardef\gamma="710D
\mathchardef\delta="710E
\mathchardef\epsilon="710F
\mathchardef\zeta="7110
\mathchardef\eta="7111
\mathchardef\theta="7112
\mathchardef\kappa="7114
\mathchardef\lambda="7115
\mathchardef\mu="7116
\mathchardef\nu="7117
\mathchardef\xi="7118
\mathchardef\pi="7119
\mathchardef\rho="711A
\mathchardef\sigma="711B
\mathchardef\tau="711C
\mathchardef\phi="711E
\mathchardef\chi="711F
\mathchardef\psi="7120
\mathchardef\omega="7121
\mathchardef\varepsilon="7122
\mathchardef\vartheta="7123
\mathchardef\varrho="7125
\mathchardef\varphi="7127

\draft

\twocolumn[\hsize\textwidth\columnwidth\hsize\csname  
@twocolumnfalse\endcsname
\title{Phase Diagram of Doped Manganates}

\author{Daniel P. Arovas$^1$ and Francisco Guinea$^{1,2}$}
\address{$^1$Department of Physics, University of California at San Diego,
 La Jolla CA 92093\\
$^2$Instituto de Ciencia de Materiales, Consejo Superior de Investigaciones
Cient{\'\i}ficas, Cantoblanco, 28049 Madrid, Spain}

\date{\today}

\maketitle

\begin{abstract}
The phase diagram of doped manganate compounds La$_{1-x}$A$_x$MnO$_3$
(with divalent A) is studied.  We analyze an extension of the double
exchange model using the Schwinger boson formalism.  Earlier work by
de Gennes on the existence of a canted phase is reproduced,
although this phase is shown to be unstable towards phase separation
in a broad regime of physical interest.  We numerically solve the mean
field equations for our model and exhibit its phase diagrams.
\end{abstract}

\pacs{PACS numbers:}
\vskip2pc]

\narrowtext

\section{Introduction}
Doped manganese oxides show many unusual features, the most
striking being the colossal magnetoresistance in the ferromagnetic
phase\cite{wk55,g63,cvm97}. The phase diagram, as function of
doping and temperature is far from elucidated. At small dopings,
many experiments are interpreted in terms of the phase diagram
proposed by de Gennes\cite{g60}, who studied the so called double
exchange model\cite{z51} (see below). Some experiments indeed
confirm the predictions derived from this approach\cite{kkk96}.
Others, however, seem to imply a more complex behavior, including
charge ordering\cite{amk96,yhn96} or coexisting phases\cite{leb96,vgf97}.
In addition, these materials show a metal-insulator transition
at low dopings and low temperatures\cite{uma95}.

In the following, we will analyze the phase diagram of these systems
using the Schwinger boson representation\cite{aa88} for the magnetic
moments, which are described by the double exchange model.
We neglect the role of lattice distortions, which may be
important at high dopings, where the Jahn-Teller
distortion present in undoped systems disappears\cite{mls95}.
The scheme that we use allows us to obtain a description of the spin waves.
In more conventional systems, it has been shown that quantum and
thermal fluctuations are adequately described\cite{aa88,sjkm89}
in this approach.  The method has already been used to study the
quasiparticle coherence in the manganese oxides\cite{s96}. Finally, the
calculations reported here are in general agreement with the work of
de Gennes\cite{g60} in the in zero temperature, large S limit.
The general features of the model are described
in the next section.  Sections III and IV adapt the Schwinger boson method
to the double exchange model.  The results for $T = 0$ are presented
in section V, where the relation of our work to the original
calculation by de Gennes\cite{g60} is discussed.  Section VI is
devoted to finite temperature results.  Finally, section VII contains
a discussion of experimental results and related thoretical work.

\section{Model}
In materials such as La$_{1-x}$A$_x$MnO$_3$\ (with A divalent),
a fraction $x$ of Mn ions are
in $3d^4$ (Mn$^{3+}$) configurations, with the remaining fraction $1-x$
in $3d^3$ (Mn$^{4+}$) states.  In a cubic crystal field, the Mn $3d$ levels
split into a lower $t_{2g}$\ triplet and an upper $e_g$\ doublet.  Intra-atomic
(``Hund's rules'') couplings overwhelm the crystal field splitting, hence the
$t_{2g}$\ levels are always triply occupied and form a
$S={\textstyle{3\over 2}}$ `core spin'.  In Mn$^{3+}$\ ions,
the $e_g$\ orbitals are further split by a static Jahn-Teller
(JT) distortion, which, together with the Hund's rules, completely determines
both the orbital as well as spin state of the $e_g$\ electron.  The $e_g$\ 
electrons may be represented by spinless, single-orbital fermions whose
hopping is modulated by the overlap of the core spin wavefunctions.
If we treat the core spin on site $i$ using the Schwinger representation,
${\mib S}^{\vphantom{\dagger}}_i={\textstyle{1\over 2}} b^\dagger_{i\alpha}
{\mib\sigma}^{\vphantom{\dagger}}_{\alpha\beta}
b^{\vphantom{\dagger}}_{i\beta}$ ($\alpha, \beta=\uparrow,\downarrow$,
$\sum_\alpha b^\dagger_{i\alpha} b^{\vphantom{\dagger}}_{i\alpha}=2S$),
then the $e_g$. electron creation operator $\psi^\dagger_{i\sigma}$ may be
factored into a spinless fermion $c$ and the Schwinger boson
$b^{\vphantom{\dagger}}_{i\sigma}$ which supplies the core spin orientation:
$\psi^\dagger_{i\sigma}=c^\dagger_i b^{\vphantom{\dagger}}_{i\sigma}$.
The role of the core spin overlap to electron hopping in these materials
is widely appreciated (see {\it e.g.\/}\ refs. \cite{g60,mls95,s96,v96}).

Neighboring core spins are coupled via superexchange through the O $2p$
orbitals \cite{g63,m97}.  For pure LaMnO$_3$ ($x=0$), the $c$-axis exchange
is antiferromagnetic whilst the exchange between neighboring ions in a
plane perpendicular to ${\hat{\mib c}}$ is ferromagnetic.  We have therefore
chosen to study the model defined by the Hamiltonian \cite{foot1}
\begin{displaymath}
{\cal H}=-{\textstyle{1\over 2}} S\sum_{{\langle ij\rangle},\sigma}
\left[ t^{\vphantom{\dagger}}_{ij} \, c^\dagger_i c^{\vphantom{\dagger}}_j \,
b^{\vphantom{\dagger}}_{i\sigma}b^\dagger_{j\sigma}+\hbox{\rm H.c.}\right]
- \sum_{\langle ij\rangle} J_{ij}\,{\mib S}_i\cdot{\mib S}_j
\end{displaymath}
for a cubic lattice of Mn ions, where the exchange $J_{ij}=-J_{\rm v}<0$ along
vertical links and $J_{ij}=J_{\rm h}>0$ along horizontal links.  Note that
this model is somewhat unrealistic in that the exchange between core spins
is fixed independent of the fermion occupancy and hence it cannot reflect
the difference between Mn$^{3+}$-Mn$^{3+}$, Mn$^{3+}$-Mn$^{4+}$,
and Mn$^{4+}$-Mn$^{4+}$\ exchange (note that Mn$^{4+}$-Mn$^{4+}$\ exchange,
appropriate to pure CaMnO$_3$, is always
antiferromagnetic \cite{m97}).  In addition, we assume a strong static
JT distortion which renders the conduction orbital nondegenerate, whereas
in the real materials this distortion vanishes for
$x\,{\raise.3ex\hbox{$>$\kern-.75em\lower1ex\hbox{$\sim$}}}\,0.2$.
Nonetheless, the model does capture what is perhaps
the most important aspect of the interaction between fermions and core
spins, namely that ferromagnetic core spin alignment leads to a larger
fermion bandwidth and reduced kinetic energy.  We aim to apply it
in the small $x$ region of the phase diagram, where most of the links are
between Mn$^{3+}$\ ions.  This hopping Hamiltonian itself, in the absence of
Heisenberg exchange terms, was considered by Sarker \cite{s96}, who
found a finite temperature transition between a ferromagnetic metal
and a spin-disordered state, presumably insulating, in which the
fermion band is completely incoherent.

\section{Mean Field Theory}
Following Sarker \cite{s96}, we invoke a Hartree-Fock decoupling of
the hopping term:
\begin{displaymath}
c^\dagger_i c^{\vphantom{\dagger}}_j{\cal F}_{ij}\to -\langle c^\dagger_i
c^{\vphantom{\dagger}}_j\rangle \langle{\cal F}_{ij}\rangle
+c^\dagger_i c^{\vphantom{\dagger}}_j\langle{\cal F}_{ij}\rangle +
\langle c^\dagger_i c^{\vphantom{\dagger}}_j\rangle{\cal F}_{ij}
+\hbox{\rm flucts.}\ ,
\end{displaymath}
where ${\cal F}_{ij}=b^{\vphantom{\dagger}}_{j\sigma}
b^\dagger_{i\sigma}$ accounts for the overlap of the core spin wave functions.
The Heisenberg exchange is treated using the static Schwinger boson mean
field theory of ref. \cite{aa88}.  The Hartree-Fock Hamiltonian is
then given by
\begin{displaymath}
{\cal H}_{\scriptscriptstyle\rm HF}={\cal H}_0+{\cal H}_{\rm hop}+
{\cal H}_{\rm Bose}+{\cal H}_{\rm cond}
\end{displaymath}
with
\begin{eqnarray}
{\cal H}_0&=&NS^2J_{\rm v}+2NS(S+1)J_{\rm h}-2S\sum_i
\Lambda^{\vphantom{\dagger}}_i\nonumber\\
&&+{2\over J_{\rm h}}\mathop{{\sum}^{\rm h}}_{\langle ij\rangle}
|Q_{\rm h}(ij)|^2+{2\over J_{\rm v}}\mathop{{\sum}^{\rm v}}_{\langle ij\rangle}
|Q_{\rm v}(ij)|^2\nonumber\\
&&+{\textstyle{1\over 2}} S\sum_{\langle ij\rangle}
\left[t_{ij}\,\langle c^\dagger_i
c^{\vphantom{\dagger}}_j\rangle\,\langle{\cal F}_{ij}\rangle
+\hbox{\rm H.c.}\right] \nonumber\\
{\cal H}_{\rm hop}&=&-{\textstyle{1\over 2}} S\sum_{\langle ij\rangle}
\left[t_{ij}\,c^\dagger_i
c^{\vphantom{\dagger}}_j\,\langle{\cal F}_{ij}\rangle+\hbox{\rm H.c.}\right]
\nonumber\\
{\cal H}_{\rm Bose}&=&\mathop{{\sum}^{\rm h}}_{\langle ij\rangle}
[Q_{\rm h}(ij)\,{\cal F}_{ij}+\hbox{\rm H.c.}]
+\mathop{{\sum}^{\rm v}}_{\langle ij\rangle}
[Q_{\rm v}(ij)\,{\cal A}_{ij}+\hbox{\rm H.c.}]\nonumber\\
&&+\sum_{i,\sigma}\Lambda^{\vphantom{\dagger}}_i\, b^\dagger_{i\sigma}
b^{\vphantom{\dagger}}_{i\sigma}
-{\textstyle{1\over 2}} S\sum_{\langle ij\rangle} [t_{ij}\,
\langle c^\dagger_i c^{\vphantom{\dagger}}_j\rangle\,{\cal F}_{ij}
+\hbox{\rm H.c.}]\nonumber\\
{\cal H}_{\rm cond}&=&-\sqrt{N}\sum_{{\mib k},\sigma}
(B^*_{{\mib k}\sigma} b^{\vphantom{\dagger}}_{{\mib k}\sigma}
+ B^{\vphantom{\dagger}}_{{\mib k}\sigma}b^\dagger_{{\mib k}\sigma})
\ .\nonumber
\end{eqnarray}
Here, $\mathop{{\sum}^{\rm h}}_{\langle ij\rangle}$ and
$\mathop{{\sum}^{\rm v}}_{\langle ij\rangle}$ represent sums over horizontal and
vertical links, respectively.  The $\Lambda^{\vphantom{\dagger}}_i$
are Lagrange multipliers which enforce the local constraints
$b^\dagger_{i\alpha}b^{\vphantom{\dagger}}_{i\alpha}=2S$,
${\cal A}_{ij}\equiv (b^{\vphantom{\dagger}}_{i\uparrow}
b^{\vphantom{\dagger}}_{j\downarrow}-b^{\vphantom{\dagger}}_{j\uparrow}
b^{\vphantom{\dagger}}_{i\downarrow})$ measures
the antiferromagnetic correlation between sites $i$ and $j$, $N$
is the total number of sites, and $B_{{\mib k}\sigma}$ is a field which
is conjugate to the Schwinger boson condensate order parameter:
\begin{displaymath}
\Psi_{{\mib k}\sigma}\equiv
{1\over\sqrt{N}}\langle b^\dagger_{{\mib k}\sigma} \rangle=-{1\over N}
\biggl\langle{\partial F\over\partial B_{{\mib k}\sigma}}\biggr\rangle
\end{displaymath}
where $F$ is the free energy.
We are guided to a simple mean field theory with seven parameters, in which
we assume
\begin{eqnarray*}
\Lambda&\equiv&\Lambda^{\vphantom{\dagger}}_i\\
{\cal F}^{\vphantom{\dagger}}_{{\rm h},{\rm v}}&\equiv&\langle
b^\dagger_{i\sigma} b^{\vphantom{\dagger}}_{j\sigma}
\rangle^{\vphantom{\dagger}}_{{\rm h},{\rm v}}\\
K^{\vphantom{\dagger}}_{{\rm h},{\rm v}}&\equiv&{\textstyle{1\over 2}}
t^{\vphantom{\dagger}}_{{\rm h},{\rm v}}S
\,\langle c^\dagger_i c^{\vphantom{\dagger}}_j
\rangle^{\vphantom{\dagger}}_{{\rm h},{\rm v}}\\
Q_{\rm h}&\equiv& Q(ij)\qquad\hbox{\rm on horizontal links}\\
Q_{\rm v}&\equiv& e^{i{\mib\pi}\cdot{\mib R}_i}\,Q(ij)
\qquad\hbox{\rm on vertical links}
\end{eqnarray*}
are all real constants, where ${\mib\pi}\equiv (0,0,\pi)$ in units where the
lattice constant is unity.  Thus, there are seven mean field parameters.

Diagonalizing ${\cal H}_{\rm Bose}$, we find
\begin{eqnarray}
{\cal H}_{\rm Bose}&=&\sum_{{\mib k},\sigma}
E({\mib k})\,\beta^\dagger_{{\mib k}\sigma}
\beta^{\vphantom{\dagger}}_{{\mib k}\sigma}+\sum_{\mib k}
\left(\sqrt{\Lambda_{\mib k}^2-\Delta_{\mib k}^2}-\Lambda_{\mib k}\right)\\
&&-N\sum_{\mib k} \pmatrix{B^*_{{\mib\pi}-{\mib k}\uparrow}&
B_{{\mib k}\downarrow}\cr}
M^{-1}({\mib k})\pmatrix{B_{{\mib\pi}-{\mib k}\uparrow}\cr
B^*_{{\mib k}\downarrow}}\nonumber
\end{eqnarray}
with
\begin{eqnarray*}
M({\mib k})=\pmatrix{\Lambda_{\mib k}-\Omega_{\mib k}&\Delta_{\mib k}\cr
\Delta_{\mib k} &\Lambda_{\mib k}+\Omega_{\mib k}}
\end{eqnarray*}
and
\begin{eqnarray*}
\Lambda_{\mib k}&=&\Lambda-2(Q_{\rm h}+K_{\rm h})(\cos k_x + \cos k_y)\\
\Delta_{\mib k}&=&-2Q_{\rm v}\cos k_z\\
\Omega_{\mib k}&=&-2K_{\rm v}\cos k_z\\
E({\mib k})&\equiv&\sqrt{\Lambda_{\mib k}^2
-\Delta_{\mib k}^2}+\Omega^{\vphantom{\dagger}}_{\mib k}\ .
\end{eqnarray*}
When there is a condensate ($T<T_{\rm c}$) the Bose spectrum is gapless, with
$\Lambda=\Lambda^*$, where
\begin{displaymath}
\Lambda^*\equiv 4(Q_{\rm h}+K_{\rm h})+2\sqrt{Q_{\rm v}^2+K_{\rm v}^2}\ .
\end{displaymath}
The dispersion then may be compared in the $x\to 0$ limit with the
spin wave result
\begin{displaymath}
E^{\rm sw}({\mib k})=S\sqrt{[J_{\rm v}+J_{\rm h}(2-\cos k_x -\cos k_y)]^2-
J_{\rm v}^2\cos^2 k_z}\ ;
\end{displaymath}
obtained by expanding about a $(0,0,\pi)$ N{\'e}el state (alternating
ferromagnetic planes).  The basic functional dependence on ${\mib k}$ is
reproduced -- this is a good preliminary check on the mean field
{\it Ansatz\/}.

Note also the particle hole symmetry present in our mean field theory.
This guarantees an $x\to 1-x$ symmetry in the phase diagram.  As mentioned
above, exchange in pure CaMnO$_3$ is different than in pure LaMnO$_3$, 
due to the presence of the second set of empty $e_g$\ states. Hence, 
this symmetry is an artifact of our model.

In deriving the mean field equations for our model, we must also include
the condensate.  The relationship between the field $B_{{\mib k}\sigma}$
and the order parameter $\Psi_{{\mib k}\sigma}$ is
\begin{equation}
\pmatrix{\Psi_{{\mib\pi}-{\mib k}\uparrow}\cr\Psi^*_{{\mib k}\downarrow}
\cr}=M^{-1}({\mib k})
\pmatrix{B_{{\mib\pi}-{\mib k}\uparrow}\cr B^*_{{\mib k}\downarrow}\cr}\ .
\end{equation}
Differentiating the condensate contribution to the free energy with
respect to a generic mean field parameter $\xi$ gives
\begin{displaymath}
{\partial F_{\rm cond}\over\partial \xi}=N\sum_{\mib k}
\pmatrix{\Psi^*_{{\mib\pi}-{\mib k}\uparrow}&\Psi^{\vphantom{\dagger}}_
{{\mib k}\downarrow}}
{\partial M({\mib k})\over\partial \xi}
\pmatrix{\Psi_{{\mib\pi}-{\mib k}\uparrow}\cr\Psi^*_{{\mib k}\downarrow}
\cr}\ .
\end{displaymath}
Enacting a global SU(2) rotation $b^{\vphantom{\dagger}}_{i\sigma}\to
U^{\vphantom{\dagger}}_{\sigma\sigma'} b^{\vphantom{\dagger}}_{i\sigma'}$,
it is easy to show that the free energy is invariant
under such a transformation.  This approach to Schwinger boson condensation
can also be applied to the cases of the uniform ferro- or antiferromagnet.
It has the comforting feature of making the SU(2)-invariance manifest from
the outset (compare {\it e.g.\/}\ with ref. \cite{sjkm89}, in which the
condensate always results in a moment in the $x$-direction).

Proceeding in our analysis, we assume condensation only at ${\mib k}=0$ and
${\mib k}=\pi$.  In order that the condensate give no contribution to the
free energy, we require that
\begin{eqnarray}
\pmatrix{\Psi^{\vphantom{\dagger}}_{{\mib\pi}\uparrow}\cr
\Psi^*_{0\downarrow}\cr}&=&-X\,e^{i\gamma}
\pmatrix{\cos{\textstyle{1\over 2}}\vartheta\cr
-\sin{\textstyle{1\over 2}}\vartheta\cr}\nonumber\\
\pmatrix{-\Psi^*_{{\mib\pi}\downarrow}\cr
\Psi^{\vphantom{\dagger}}_{0\uparrow}\cr}&=&-Y\,e^{i\delta}
\pmatrix{\cos{\textstyle{1\over 2}}\vartheta\cr
-\sin{\textstyle{1\over 2}}\vartheta\cr}
\label{XYeqn}
\end{eqnarray}
where $\tan\vartheta=Q_{\rm v}/K_{\rm v}$ and $X$, $Y$, $\gamma$,
and $\delta$ are
at this point arbitrary parameters specifying the direction and magnitude
of what is in general a canted $(0,0,\pi)$ antiferromagnet \cite{g60}.
Equation \ref{XYeqn} also is consistent with the free energy being a
convex function of the order parameter $\Psi_{{\mib k}\sigma}$.
The condensate is then spatially varying, with
\begin{equation}
\langle b^{\vphantom{\dagger}}_{i\sigma}\rangle=\Psi_{0\sigma}
+ \Psi_{{\mib\pi}\sigma}\, e^{i{\mib\pi}\cdot{\mib R}_i}\ .
\end{equation}
The condensate contribution to the local magnetization
$\langle{\mib S}_i\rangle$
is easily computed and $\vartheta$ is found to be the canting angle.

\section{Mean Field Equations}
We now are in a position to write down the mean field equations.
We work in the grand canonical ensemble, introducing a chemical potential
$\mu$ for the fermions.  This introduces an eighth parameter.
However, we find that the mean field equations guarantee
$Q_{\rm h}={\textstyle{1\over 2}}J_{\rm h}{\cal F}_{\rm h}$
always, so we are left with the following seven equations:
\begin{eqnarray}
2S&=&\int\!\!{d^3\!k\over (2\pi)^3} {\Lambda^{\vphantom{\dagger}}_{\mib k}
\over\sqrt{\Lambda_{\mib k}^2-\Delta_{\mib k}^2}}\,
{\rm ctnh\,}\Bigl({E({\mib k})
\over 2{k_{\scriptscriptstyle\rm B}T}}\Bigr) + R^2-1\nonumber\\
{Q_{\rm v}\over J_{\rm v}}&=&\int\!\!{d^3\!k\over (2\pi)^3}
{Q_{\rm v}\cos^2 k_z\over\sqrt{\Lambda_{\mib k}^2-\Delta_{\mib k}^2}}\,
{\rm ctnh\,}\Bigl({E({\mib k})
\over 2{k_{\scriptscriptstyle\rm B}T}}\Bigr) +
{Q_{\rm v} R^2\over 2\sqrt{Q_{\rm v}^2+K_{\rm v}^2}}\nonumber\\
{\cal F}_{\rm h}&=&\int\!\!{d^3\!k\over (2\pi)^3}
{(\cos k_x + \cos k_y)\,\Lambda^{\vphantom{\dagger}}_{\mib k}\over
2\sqrt{\Lambda_{\mib k}^2-\Delta_{\mib k}^2}}\,
{\rm ctnh\,}\Bigl({E({\mib k})
\over 2{k_{\scriptscriptstyle\rm B}T}}\Bigr) + R^2\nonumber\\
{\cal F}_{\rm v}&=&2\int\!\!{d^3\!k\over (2\pi)^3}
{\cos k_z\over\exp\Bigl({E({\mib k})
\over{k_{\scriptscriptstyle\rm B}T}}\Bigr)-1}\,
+{K_{\rm v} R^2\over \sqrt{Q_{\rm v}^2+K_{\rm v}^2}}\nonumber\\
K_{\rm h}&=&{\textstyle{1\over 4}} St_{\rm h}\int\!\!{d^3\!k\over (2\pi)^3}
{\cos k_x + \cos k_y\over
\exp\Bigl({\epsilon({\mib k})-\mu
\over{k_{\scriptscriptstyle\rm B}T}}\Bigr)+1}\nonumber\\
K_{\rm v}&=&{\textstyle{1\over 2}} St_{\rm v}\int\!\!{d^3\!k\over (2\pi)^3}
{\cos k_z\over
\exp\Bigl({\epsilon({\mib k})-
\mu\over{k_{\scriptscriptstyle\rm B}T}}\Bigr)+1}\nonumber\\
x&=&\int\!\!{d^3\!k\over (2\pi)^3}{1\over\exp
\Bigl({\epsilon({\mib k})-
\mu\over{k_{\scriptscriptstyle\rm B}T}}\Bigr)+1}
\end{eqnarray}
where $R=\sqrt{X^2+Y^2}$ is the condensate amplitude, $x$ is the hole
concentration, and
\begin{displaymath}
\epsilon({\mib k})=-St_{\rm h}{\cal F}_{\rm h}
(\cos k_x + \cos k_y) -St_{\rm v}{\cal F}_{\rm v}\cos k_z
\end{displaymath}
is the fermion dispersion.  The integrals are performed over the first
Brillouin zone of the cubic lattice.  There are seven mean field
equations corresponding to seven mean field parameters.  The parameters are:
\begin{eqnarray*}
T<T_{\rm c}&:\quad&Q_{\rm v},{\cal F}_{\rm h},{\cal F}_{\rm v},
K_{\rm h},K_{\rm v},\mu,R\quad(\Lambda=\Lambda^*)\\
T>T_{\rm c}&:\quad&Q_{\rm v},{\cal F}_{\rm h},{\cal F}_{\rm v},
K_{\rm h},K_{\rm v},\mu,\Lambda\quad(R=0)\ .
\end{eqnarray*}

We have identified several phases which emerge from the mean field theory:
\begin{eqnarray*}
\hbox{\rm (I)}&\ &\hbox{\rm Antiferromagnet (LRO at ${\mib\pi}$)}:
{\cal F}_{\rm v},K_{\rm v}=0, R\neq 0\\
\hbox{\rm (II)}&\ &\hbox{\rm Canted (LRO at $0$ and ${\mib\pi}$)}:
Q_{\rm v},K_{\rm v},R\neq 0\\
\hbox{\rm (III)}&\ &\hbox{\rm Ferromagnet (LRO at $0$)}:
Q_{\rm v}=0, R\neq 0\\
\hbox{\rm (IV)}&\ &\hbox{\rm 3d Local Magnetic Order}: {\cal F}_{\rm v},
{\cal F}_{\rm h}\neq 0, R=0\\
\hbox{\rm (V)}&\ &\hbox{\rm 2d Local Magnetic Order}: {\cal F}_{\rm v},R=0,
{\cal F}_{\rm h}\neq 0\\
\hbox{\rm (VI)}&\ &\hbox{\rm Maximally Disordered}: Q_{\rm v},{\cal F}_{\rm v},
{\cal F}_{\rm h},K_{\rm v},K_{\rm h},R=0\ .
\end{eqnarray*}
The ordered phases I, II, and III were identified by deGennes
\cite{g60}; we have also found evidence of phase separation below $T_{\rm c}$
(see also \cite{n96,dym97}).  In what follows we describe our analytical and
numerical investigations of the phase diagram.

\section{$T=0$, $S\to\infty$ Limit}
Our mean field equations simplify considerably in the limit of zero
temperature and $S\to\infty$.  We examine the three ordered
($\Lambda=\Lambda^*$, $R>0$) phases,
\begin{eqnarray*}
\hbox{\rm (I)}&\ &Q_{\rm v}=SJ_{\rm v},\ {\cal F}_{\rm h}=2S,
\ {\cal F}_{\rm v}=K_{\rm v}=0\\
\hbox{\rm (II)}&\ &Q_{\rm v}^2+K_{\rm v}^2=S^2J_{\rm v}^2,
\ {\cal F}_{\rm v}=2K_{\rm v}/J_{\rm v},\ {\cal F}_{\rm h}=2S\\
\hbox{\rm (III)}&\ &Q_{\rm v}=0,\ {\cal F}_{\rm v}={\cal F}_{\rm h}=2S\ .
\end{eqnarray*}
The canted phase II can smoothly interpolate between the
${\mib\pi}$-antiferromagnet
I and the ferromagnet III, with $\vartheta$ going from
${\textstyle{1\over 2}}\pi$ to $0$.
We start with the canted structure, solving the mean field equation
${\cal F}_{\rm v}=2K_{\rm v}/J_{\rm v}$.  We do this in the regime
$x\ll 1$ by expanding the fermion
dispersion relation 
\begin{displaymath}
\epsilon({\mib k})=\epsilon(0)+St_{\rm v}
{\cal F}_{\rm v}(1-\cos k^{\vphantom{\dagger}}_z)+{\textstyle{1\over 2}}
S t_{\rm h}{\cal F}_{\rm h}(k_x^2+k_y^2)+\ldots\ ;
\end{displaymath}
since ${\cal F}_{\rm v}=0$ is a possible solution, we keep the full
$c$-axis dispersion. We find that the solution is characterized by the
dimensionless parameter $r\equiv 8\pi J_{\rm v}t_{\rm h}/t_{\rm v}^2$.
For $r>1$, the only solution has ${\cal F}_{\rm v}=K_{\rm v}=0$,
and we have antiferromagnetism at finite doping.  However, experiments
\cite{hmr97} suggest $J_{\rm v}\approx 0.58\,$meV while spin density functional
calculations \cite{ps96} suggest $t_{\rm v}\approx 44\,$meV (the physical
hopping parameter is $S^2 t\approx 100\,$meV).  This gives $r\approx 0.33$,
so the $r>1$ regime is unphysical for La$_{1-x}$A$_x$MnO$_3$.
The ground state energy per site is found to be
\begin{equation}
{E^{\vphantom{\dagger}}_{\rm I}\over NS^2}=-2J_{\rm h}-J_{\rm v}-4t_{\rm h} x
+2\pi t_{\rm h} x^2 + \ldots
\end{equation}

For $r<1$, we have a solution with nonzero ${\cal F}_{\rm v}$.  We find
\begin{eqnarray}
{E^{\vphantom{\dagger}}_{\rm II}\over NS^2}&=&-2J_{\rm h}-J_{\rm v}-
4t_{\rm h} x +2\pi g(r) t_{\rm h} x^2 + \ldots\\\
g(r)&=&{\pi^2 r -\pi\alpha\sin^2\alpha\over (\sin\alpha-\alpha\cos\alpha)^2}
\nonumber
\end{eqnarray}
where $\alpha(r)\in[0,\pi]$ is defined implicitly by the equation
\begin{equation}
\alpha-\sin\alpha\cos\alpha=\pi r\ .
\end{equation}
We obtain ${\cal F}_{\rm v}=xt_{\rm v}/J_{\rm v}+{\cal O}(x^2)$ in this regime.
For $r<{\textstyle{1\over 2}}$, $g(r)<0$ and the coefficient of
$x^2$ is negative, indicating phase separation.  
For ${\textstyle{1\over 2}} < r < 1$, $g(r)>0$ and
there is a homogeneous, thermodynamically stable canted phase,
originally identified by deGennes \cite{g60}.  Our estimate $r\approx 0.33$
suggests that phase separation is likely.

Within the canted phase, as $x$ increases from $0$ to ${\textstyle{1\over 2}}$, 
the canting angle decreases from $\vartheta={\textstyle{1\over 2}}\pi$
(${\mib\pi}$-LRO) to
$\vartheta=0$ ($0$-LRO).  
The transition from canted to ferromagnetic
order is continuous, occuring at a critical concentration $x^*$ 
determined by the simultaneous solution of the two equations
\begin{eqnarray}
x^*&=&\int\!\!{d^3\!k\over (2\pi)^3} {\rm\Theta}
(\mu-\epsilon({\mib k}))\nonumber\\
{2J_{\rm v}\over t_{\rm v}}&=&\int\!\!{d^3\!k\over (2\pi)^3} \cos k_z\,
{\rm\Theta}(\mu-\epsilon({\mib k}))
\end{eqnarray}
in the two variables $x^*$ and $\mu$.  For sufficiently large $J_{\rm v}$, 
$\vartheta$ is nonzero even at half-filling and the system remains canted
for all $x$.

\begin{figure} [t]
\centering
\leavevmode
\epsfxsize=8cm
\epsfysize=8cm
\epsfbox[18 144 592 718] {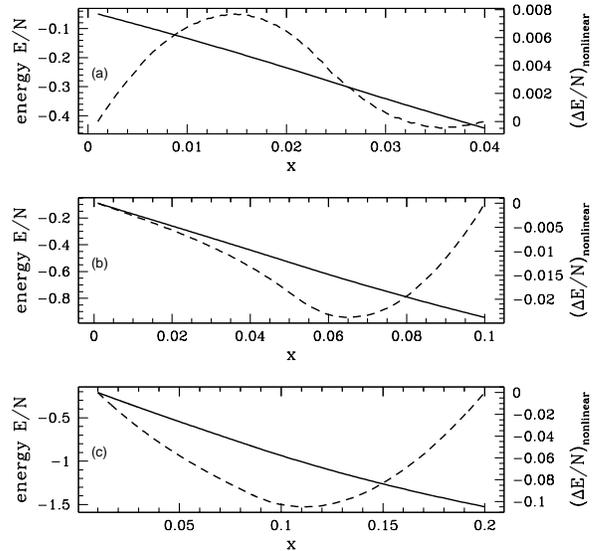}
\caption[]
{\label{fig1} Energy per site $E/N$ (solid) {\it versus\/} concentration $x$
for $J_{\rm v}=J_{\rm h}\equiv J$, $t_{\rm v}=t_{\rm h}\equiv t=1$
at temperature $T=0.01\,t$
for three different values of $r=8\pi J/t$: (a) $r=0.25$, (b) $r=0.50$,
(c) $r=0.75$.  To ascertain the sign of $\partial^2\!E/\partial x^2$, we have
subtracted from $E/N$ the linear part; the dashed curve is the remaining
contribution.  Note that $E(x)$ is convex for small $x$ in (a) and (b),
indicative of phase separation.}
\end{figure}

\begin{figure} [t]
\centering
\leavevmode
\epsfxsize=8cm
\epsfysize=8cm
\epsfbox[18 144 592 718] {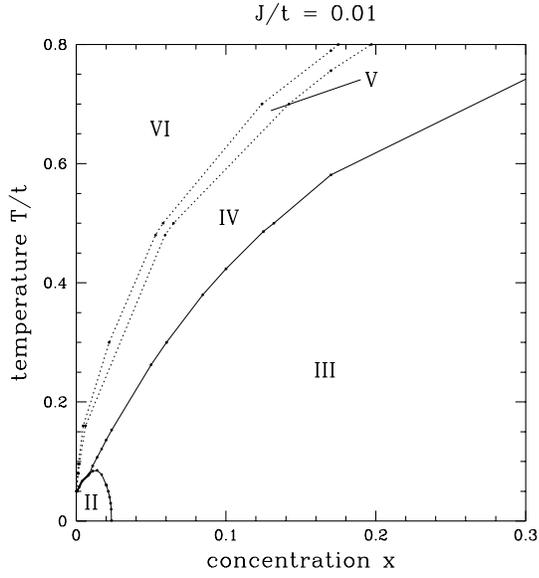}
\caption[]
{\label{fig2} Phase diagram for $J/t=0.01$ ($r=0.251$) obtained from
numerical solution of the mean field equations.  The dark solid line
separating phases II and IV is first order.  All other transitions are
second order.  Dotted lines represent transitions between disordered states.
Phase separation occurs outside region III (see text for discussion).}
\end{figure}

\begin{figure} [!t]
\centering
\leavevmode
\epsfxsize=8cm
\epsfysize=8cm
\epsfbox[18 144 592 718] {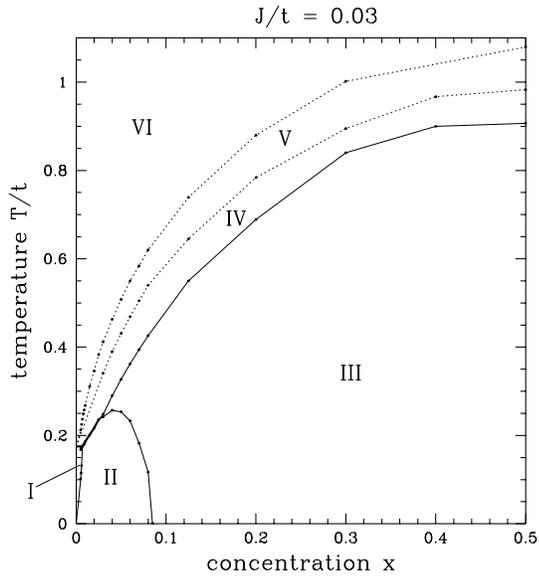}
\caption[]
{\label{fig3} Phase diagram for $J/t=0.03$ ($r=0.754$) obtained from
numerical solution of the mean field equations.  A small sliver of
the ${\mib\pi}$-ordered phase exists in the lower left corner.  The dark
solid line separating phases I and IV as well as II and IV is first order.
All other transitions are second order.  As in figure 2, dotted lines
represent transitions between disordered states.}
\end{figure}

\section{Finite $T$ Phases}
To explore the finite temperature phases of our model, we have solved
the mean field equations numerically using the {\tt MINPACK} routine
{\tt hybrd.f}.  (To simplify matters, we assumed $J_{\rm v}=J_{\rm h}\equiv J$
and $t_{\rm v}=t_{\rm h}\equiv t$.)  We found that there is
often more than one solution
to the mean field equations; in such cases we computed the energy of
each solution and identified the minimum energy state.  To compare with
the analytical results of the previous section, we computed the energy
{\it versus\/} concentration $x$ for three different values of $r$,
all at a temperature well below $T_{\rm c}$ (see figure \ref{fig1}).
This verified our prediction of phase separation for
$0<r<{\textstyle{1\over 2}}$.

In figures 2 and 3 we plot phase diagrams for $J/t=0.01$ ($r=0.251$)
and $J/t=0.03$ ($r=0.754$), respectively.  Regions are labeled I through VI
corresponding to the six phases discussed above (recall that a condensate
is present only in phases I, II, and III).  We find that only region III
of figure 2 is stable with respect to phase separation.  The homogeneous
mean field solution yields a convex $F(x,T)$ outside of this region up to
temperatures on the order of
$T_*(x)\,{\raise.3ex\hbox{$<$\kern-.75em\lower1ex\hbox{$\sim$}}}\,0.86\,t$;
for $T>T_*$ the free energy is concave in $x$.

We found a first order line separating phases I and II from the disordered
phases.  This is so even at $x=0$ -- the mean field theory predicts
a first order transition from the $(0,0,\pi)$ N{\'e}el state to a
magnetically disordered state.  This is perhaps a worrisome artifact of
the mean field theory.  In addition, the transitions between the
disordered phases may well become smooth crossovers when fluctuation
effects are accounted for.  Indeed, application of the Schwinger
boson formalism to the Heisenberg model \cite{aa88} leads to a spurious
high temperature mean field transition to a state in which the
magnon bandwidth vanishes, analogous to phase VI.

Sarker \cite{s96} has discussed the behavior of the electron spectral
function and found it to be entirely incoherent when the core spins are
disordered.  Writing (at finite temperature $T$)
\begin{eqnarray*}
G^{\rm R}_{\alpha\beta}({\mib k},t)
&=&-i\langle \{\psi^{\vphantom{\dagger}}_{{\mib k}\alpha}(t),
\psi^\dagger_{{\mib k}\beta}(0)\}\rangle\,\Theta(t)\\
G^{\rm R}_{\alpha\beta}({\mib k},\omega)&=&\int_{-\infty}^\infty\!\!\!
d\omega'\,{\rho^{\vphantom{\dagger}}_{\alpha\beta}({\mib k},\omega')\over
\omega-\omega'+i0^+}\ ,
\end{eqnarray*}
we find, at the mean field level,
\begin{eqnarray}
\rho^{\vphantom{\dagger}}_{\alpha\beta}({\mib k},\omega)&=&
\int\!\!{d^3\!q\over (2\pi)^3}
\left[ n^{\vphantom{\dagger}}_{\alpha\beta}({\mib q}) +
\delta^{\vphantom{\dagger}}_{\alpha\beta}\,f({\mib k}+{\mib q})
\right]\nonumber\\
&&\qquad\times\delta\left(\omega+E({\mib q})
-\epsilon({\mib k}+{\mib q})+\mu\right)\ ,
\end{eqnarray}
where $n^{\vphantom{\dagger}}_{\alpha\beta}({\mib q})=
\langle b^\dagger_{{\mib k}\alpha}\,b^{\vphantom{\dagger}}_{{\mib k}\beta}
\rangle$ and $f({\mib p})=\langle c^\dagger_{\mib p}\,
c^{\vphantom{\dagger}}_{\mib p}\rangle$ are equilibrium averages.
The contribution of the condensate to the spectral density results in
well-defined quasiparticle peaks.  For instance,
\begin{eqnarray*}
\rho^{\rm cond}_{\uparrow\uparrow}({\mib k},\omega)&=&
Y^2\sin^2(\vartheta/2)\,\delta(\omega+E(0)-\epsilon({\mib k})+\mu)\\
&&\ +X^2\cos^2(\vartheta/2)\,\delta(\omega+E({\mib \pi})-
\epsilon({\mib \pi}-{\mib k})+\mu)\ ,
\end{eqnarray*}
where $X$, $Y$, and $\vartheta$ describe the amplitude and orientation of
the condensate (recall equation \ref{XYeqn}).  The remaining contribution to
the spectral function, $\Delta\rho$, is incoherent and spectrally broad
\cite{s96}.  Our calculation allows for condensation both at
${\mib k}=0$ as well as ${\mib k}={\mib \pi}$, and as expected there
are two quasiparticle peaks when translational symmetry
is broken (phase I).  A detailed study of $\rho({\mib k},\omega)$
in the various ordered and disordered phases is pending.

\section{Conclusions}
We have shown that the double exchange model has a variety of
possible phase diagrams, controlled by the parameter 
$r\equiv 8\pi J_{\rm v}t_{\rm h}/t_{\rm v}^2$. For realistic values of $r$,
$0.05 \le r \le 0.2$, we find a phase diagram similar to the one 
proposed by de Gennes\cite{g60}, except that the canted phase
is replaced by a region of phase separation.
In addition, the transition to the paramagnetic phase may be of
first order, and the paramagnetic phase itself is anisotropic.
For $r > {\textstyle{1\over 2}}$, we find that the canted phase is stable.
The main source of uncertainty in the value of $r$
arises from the lack of a precise determination of the
hoppings, which may depend on the composition and details
of the lattice structure\cite{cvm97}.  Perhaps
both situations may be realized experimentally.

Coulomb interactions will prevent charge separation at large scales.
The electrostatic energy required to break the system into charged
domains of side $\ell$ goes as ${e^2 x^2/\epsilon\ell}$, where $\epsilon$
is the dielectric constant.
The magnetic energy cost to create a domain wall of size $\ell$
is roughly $J_{\rm v} (\ell/a)^2$, where $a$ is the lattice spacing.  Hence,
the domain size, $\ell$ goes as $a^{4/3}(J_{\rm v}\epsilon/e^2 x^2)^{1/3}$,
which should be on the order of a few lattice constants.

Phase separation in these compounds has previous been discussed
phenomenologically in \cite{n96}, and in \cite{dym97},
in the context of numerical results for the ferromagnetic Kondo lattice
in one, two, and infinite dimensions.  At large values
of the Hund's rule coupling, this model reduces to the double exchange model,
plus antiferromagnetic interactions between the core spins.
In-plane ferromagnetic interactions do not arise, as they
are induced by the second $e_g$\ band. Our results, for $T = 0$,
are qualitatively
in agreement with those reported in\cite{dym97}, provided that
one identifies our canted phase with the incommensurate order
reported there.  

Our calculation reproduces the observed magnon spectrum at zero
doping\cite{mhr96}, except for a small anisotropy gap.
In the canted phase, the long wavelength spin waves
behave as $\sqrt{v^{\vphantom{\dagger}}_\parallel
( k_x^2 + k_y^2 ) + v^{\vphantom{\dagger}}_\perp k_z^2}$.
In the phase separated regime, localized ferro- and antiferromagnetic modes
are expected, which may have been observed experimentally\cite{hmr97}.

A detailed study of transport properties lies beyond the scope of the present
calculation. It is interesting to note, however, that, in the presence of
phase separation, hopping is suppressed in the out of plane direction in
the antiferromagnetic domains. This effect can contribute to
make these compounds insulating at low dopings, in agreement with
experiments.

\section{Acknowledgements}
We are particularly grateful to J. M. D. Coey for educating
us on many issues regarding manganates.  We also thank A. Millis for
useful comments.  F. G. acknowledges the hospitality
of the University of California at San Diego, where this work 
was performed.


\begin{references}
\bibitem{foot1} We have scaled the hopping integrals by $S^2$ in order
to allow a large-$S$ analysis of our mean field equations.
\end{references}
\end{document}